\documentclass{article}
\usepackage{graphicx}
\usepackage{amssymb,amsmath}

\begin{document}
\baselineskip 18pt

\begin{center}
{\Large {\bf Inducing periodicity in lattices of chaotic maps with advection}} 

\vskip5mm 
Pedro G.~Lind$^{1,2}$\footnote{Email address: lind@ica1.uni-stuttgart.de} and
Jason A.C.~Gallas$^{1,2}$\footnote{Email address: jgallas@if.ufrgs.br} 
\vskip3mm \mbox{}%
$^{1}$ Institut f\"ur Computer Anwendungen, Universit\"at Stuttgart,
       Pfaffenwaldring, 27, \\ D-70569 Stuttgart, Germany

\mbox{}$^{2}$ Instituto de F\'\i sica, Universidade Federal do 
              Rio Grande do Sul,\\ 91501-970 Porto Alegre, Brazil


\bigskip

\begin{abstract}
We investigate a lattice of coupled logistic maps where, in addition to 
the usual diffusive coupling, an advection term parameterized 
by an asymmetry in the coupling is introduced.
The advection term induces periodic behavior on a 
significant number 
of non-periodic solutions of the purely diffusive case.
Our results are based on the characteristic exponents for such systems, 
namely the mean Lyapunov exponent and the co-moving Lyapunov exponent. 
In addition, we study how to deal with more complex phenomena in which
the advective velocity may vary from site to site.
In particular, we observe wave-like pulses to appear and disappear
intermittently whenever the advection is spatially inhomogeneous.

\bigskip
\noindent PACS numbers: 89.75.Kd   
                        05.45.Xt,  
                        05.45.Ra   
\end{abstract}
\end{center}
\newpage

\section*{1. Introduction}
\label{sec:intro}

Recent studies of pattern formation and pattern dynamics in spatially
extended systems have provided important clues to understand
nonlinear mechanisms of nonequilibrium conditions in many physical 
phenomena such as, e.g.,
laser dynamics~\cite{pre-lasers}, 
electroconvection~\cite{pre-electroconv},
Rayleigh-B\'enard convection~\cite{pre-hydr},
field-induced phenomena in magnetic fluids~\cite{pre-mag},
and many others\cite{cross}.
A very popular way to study pattern formation is by using networks
of coupled oscillators ruled locally by time-discrete mappings, 
the so-called coupled map lattices\cite{kanekobooks}.
Such networks of maps were applied e.g.~to 
model information coding~\cite{cml-mackey},
to study nonlinear wave-like patterns~\cite{asym},
ocean convection parameterization~\cite{pots},
synchronization processes~\cite{bocca,synclind}.

Usually, coupled map lattices are regarded as a discretize version
of reaction-diffusion systems~\cite{kanekobooks}, because they involve a set 
of discrete-time nonlinear oscillators coupled through diffusion
alone.
Pattern formation in networks of maps was already studied 
for this purely diffusive regime\cite{kanekobooks,ijbc}.
However, as is well-known\cite{landau}, spatially extended 
phenomena are quite frequently subject not only to diffusion but also to 
advection.
Denoting by $\gamma_i$ the advection strength at site $i$, 
a simple way of incorporating advection in networks of maps was proposed 
recently\cite{advec} being embodied in the following equation of motion
\begin{equation} 
    x_{t+1}(i) = f(x_t(i))+\varepsilon{\cal D}_{i,t}
                          -\gamma_i{\cal A}_{i,t}\  ,  \label{model}
\end{equation}
where $i=1,\dots,L$, with $L$ being the lattice size,
$\varepsilon$ represents the diffusion.
Following common practice, we also use the logistic map,
$f(x)=1-ax^2$, to drive the local dynamics.
Here ${\cal D}_{i,t}$ and ${\cal A}_{i,t}$ are discretized forms 
of the diffusion and advection operators,  namely
\begin{subequations}   \label{terms}
\begin{eqnarray*}
\hspace{-0.3cm}
{\cal D}_{i,t} &=& \frac{f(x_t(i+1))+f(x_t(i-1))}{2}-f(x_t(i)), 
                     \label{difus}\\
\hspace{-0.3cm}
{\cal A}_{i,t} &=& \frac{f(x_t(i+1))-f(x_t(i-1))}{2}. \label{advec}
\end{eqnarray*}
\end{subequations}
As shown in Ref.~\cite{asym}, $\gamma_i$ must be in the interval
$-\varepsilon\le\gamma_i\le\varepsilon$.
We impose periodic boundary condition: $x_t(L+1) \equiv x_t(1)$.

For $\gamma=0$ Eq.~(\ref{model}) reduces to the well-known purely diffusive
model, while for $\gamma=\pm\varepsilon$ one obtains the two possible
one-way coupling regimes\cite{kanekobooks}.
In the presence of advection one may distinguish two different situations: 
{\it homogeneous advection}, when 
$\gamma_i\equiv \gamma$  for all sites $i$,  and  
{\it inhomogeneous advection}, when the $\gamma_i$ are free to vary along 
the lattice.
In an earlier investigation\cite{advec} we described how the spatial 
periodicity (wavelength) of wave-like patterns evolve with the advection 
strength.
The purpose of this paper is to show how advection affects the {\it temporal\/} 
stability of patterns.
In particular, we address the question of how and under which conditions 
advection changes pattern evolutions from chaotic to  periodic evolutions.

The stability of solutions in dynamical systems is established by
the well-known Lyapunov analysis, considering the so-called local
Lyapunov exponents, defined from the logarithm of the eigenvalues of the 
system.
If the evolution is periodic one finds a negative exponent while for
chaotic evolutions the exponent is positive.

Local Lyapunov exponents are computed in a static frame.
However, recently~\cite{kanekobooks,boffetta01} it was shown that, for 
particular situations such as one-way coupling regimes, unstable
perturbations may travel along the lattice, with a corresponding 
{\it negative\/} local Lyapunov exponents. 
Therefore, local Lyapunov exponents may not be always suitable to characterize
the stability of such solutions. 
To cure this shortcoming one uses the so-called co-moving 
Lyapunov exponents\cite{kanekobooks,boffetta01}.

Co-moving Lyapunov exponents are defined from the eigenvalues of a
Jacobian computed in a moving frame having some `velocity'.
The most important feature of co-moving exponents lies in the
fact that they allow to discriminate between 
(i) absolute stability,
(ii) absolute instability and (iii) convective instability, which is a possible
feature in spatially extended systems.
Furthermore, with co-moving Lyapunov exponents one may study,
e.g., transitions from regular patterns to spatiotemporal 
intermittency\cite{boffetta01},
propagation of correlations~\cite{giacomelli},
and predictability\cite{ziehmann}.

In this paper we study the stability of pattern evolutions using
both local and co-moving Lyapunov exponents.
In Section 2 we start with a situation where advection is
homogeneously distributed. In Section 3 we consider nontrivial
behaviors of pattern evolutions when advection varies in space.
Final conclusions are given in Section 4.

\section*{2. Switching non-periodic into periodic evolutions}
\label{sec:lyap}

The purpose of this Section is to study the influence of diffusion 
and homogeneous advection in pattern evolutions when ruled by
Eq.~(\ref{model}) with $\gamma_i\equiv\gamma$.
We start computing  the local Lyapunov exponents, given by the
logarithm of the maximum absolute value of all eigenvalues of the
Jacobian of Eq.~(\ref{model}).

\begin{figure}[!hbp]
\begin{center}
\includegraphics[width=5.3cm]{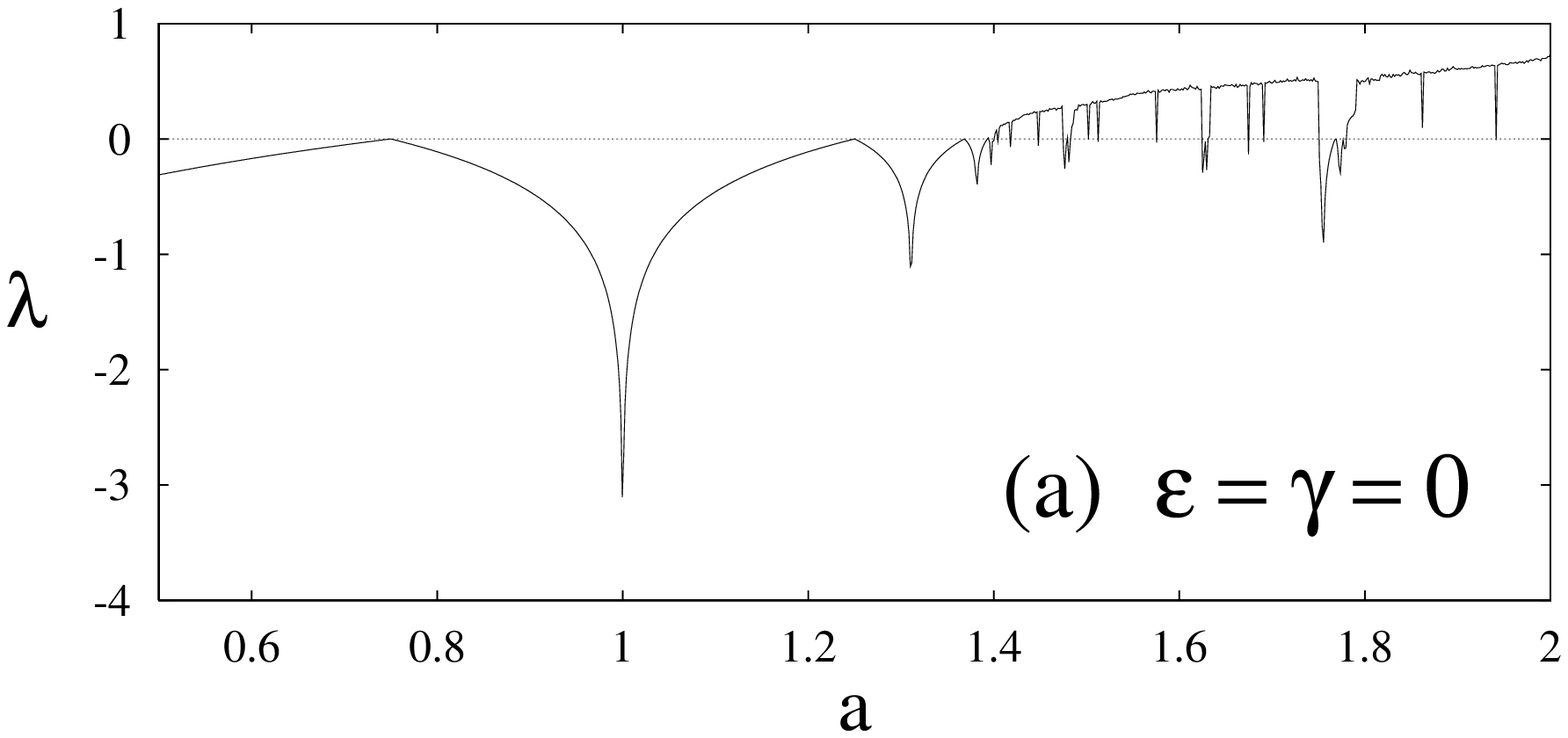}%
\includegraphics[width=5.3cm]{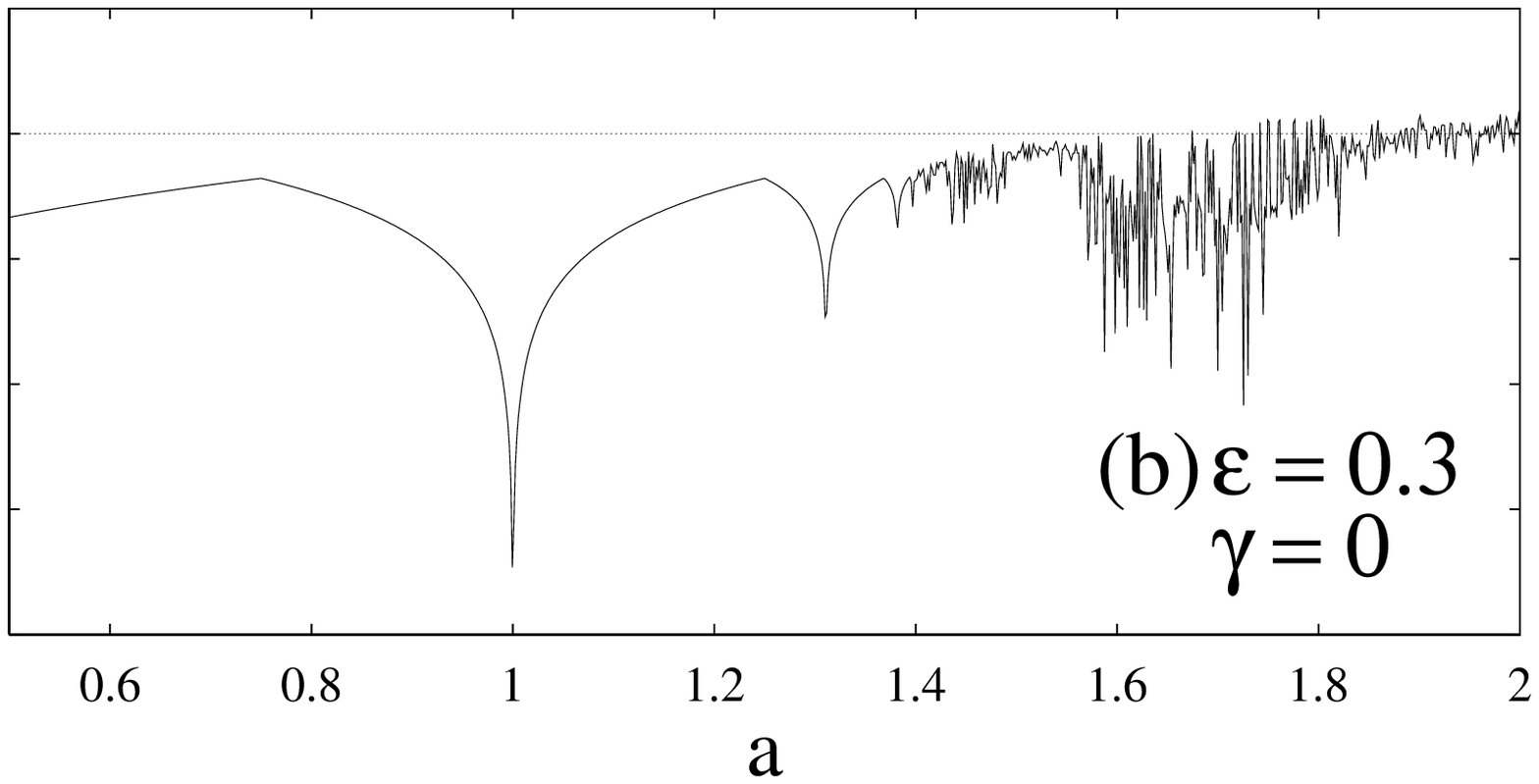}%
\includegraphics[width=5.3cm]{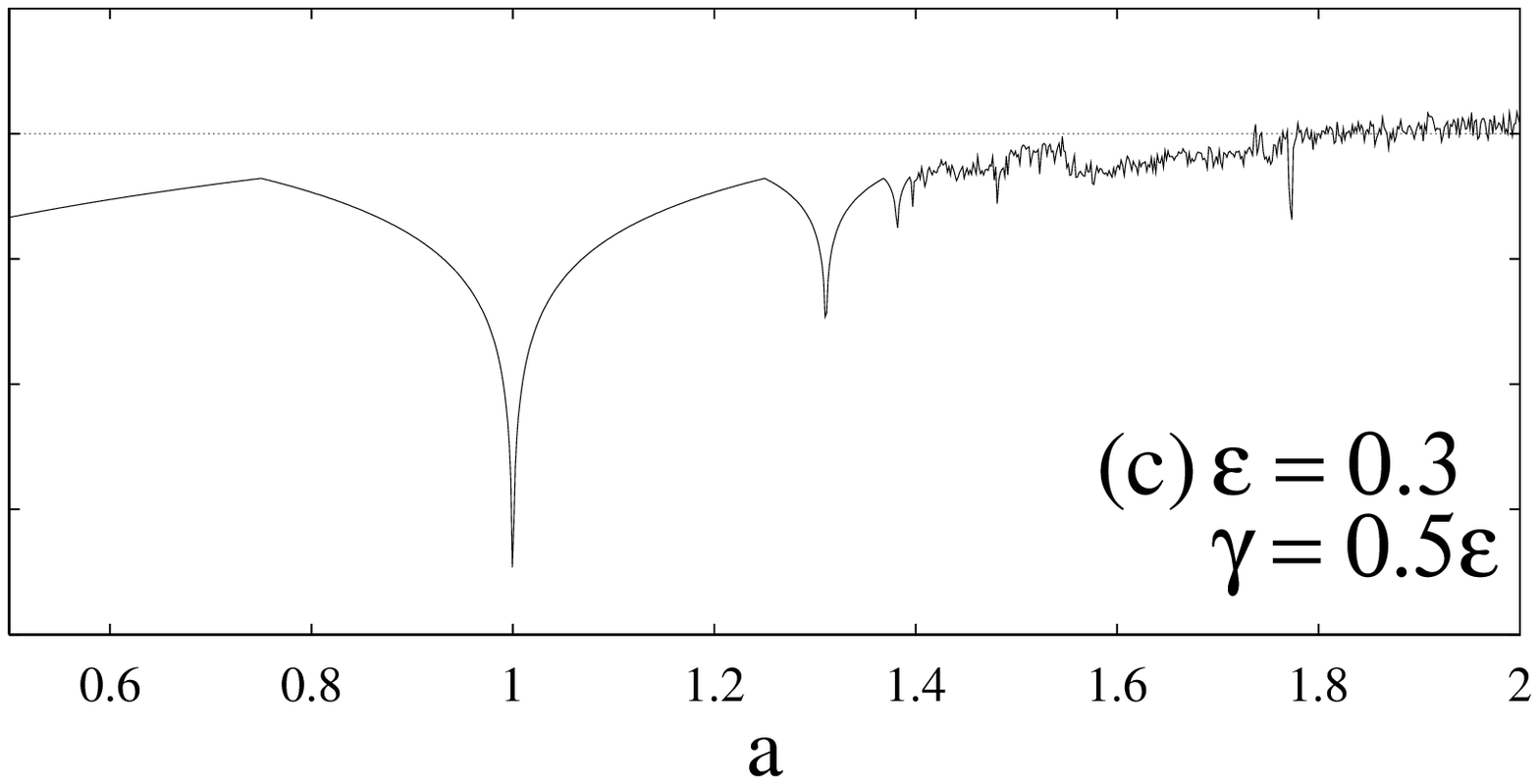}
\end{center}
\caption{\protect
  Local Lyapunov exponent $\lambda$ as a function of local nonlinearity $a$,  
  {\bf (a)} for the uncoupled regime ($\varepsilon=0$ and $\gamma=0$),
  {\bf (b)} for the purely diffusive regime ($\gamma=0$) with
  $\varepsilon=0.3$, and
  {\bf (c)} for $\varepsilon=0.3$ and $\gamma=0.5\varepsilon$.
  Here $L=64$.}
\label{fig1}
\end{figure}

For the uncoupled regime ($\varepsilon=0$ and $\gamma=0$), local Lyapunov 
exponents reduce to the Lyapunov exponents of the local map, as
illustrated in Fig.~\ref{fig1}a.
We use a sample of $100$ sets of initial conditions of the form
$x_0(i)=x^{\ast}+\delta\theta_r(i)$ where 
$x^{\ast}=-(1+\sqrt{1+4a})/2$
is the unstable fixed point of the local logistic map, 
$\theta_r(i)$
is a homogeneously distributed random value in the range $[0,1]$ and $\delta=0.001$.
The Jacobian of Eq.~(\ref{model}) is computed over $100$ time-steps
after discarding a transient of $100.000$ time-steps.
For purely diffusive coupling ($\varepsilon\neq 0$) one
knows\cite{kanekobooks}  that local Lyapunov exponents decrease.
This fact is illustrated in Fig.~\ref{fig1}b, where $\varepsilon=0.3$,
showing a Lyapunov spectrum which, before the accumulation
point $a=1.401155\dots$\cite{ijbc}, has a similar shape as that of the 
uncoupled regime.
Above the accumulation point one observes a much larger exponent fluctuation 
then that of the uncoupled regime.

Figure \ref{fig1}c shows, for the same conditions of Fig.~\ref{fig1}a 
and \ref{fig1}b, the local Lyapunov exponent as a function
of the nonlinearity when diffusion and advection are simultaneously present,
namely $\varepsilon=0.3$ and $\gamma=0.5\varepsilon$.
Comparing Figs.~\ref{fig1}b and \ref{fig1}c one concludes that,
apparently, advection does not change significantly the 
local Lyapunov spectrum.
The main difference between the presence and absence of advection
is observed above the accumulation point, where the large fluctuations
observed in the purely diffusive regime are then shortened.

In general, these observations remain valid for any diffusion and
advection strengths.
For either periodic or chaotic local dynamics\cite{kanekobooks},
our simulations have shown that diffusion promotes the 
stabilization of local amplitudes. 
Numerical fitting of the Lyapunov exponents show
the $\varepsilon$-dependence to be quadratic, 
i.e.~$\lambda\propto \sqrt{1-\varepsilon}$.

\begin{figure}[!hbp]
\begin{center}
\includegraphics[width=8.0cm]{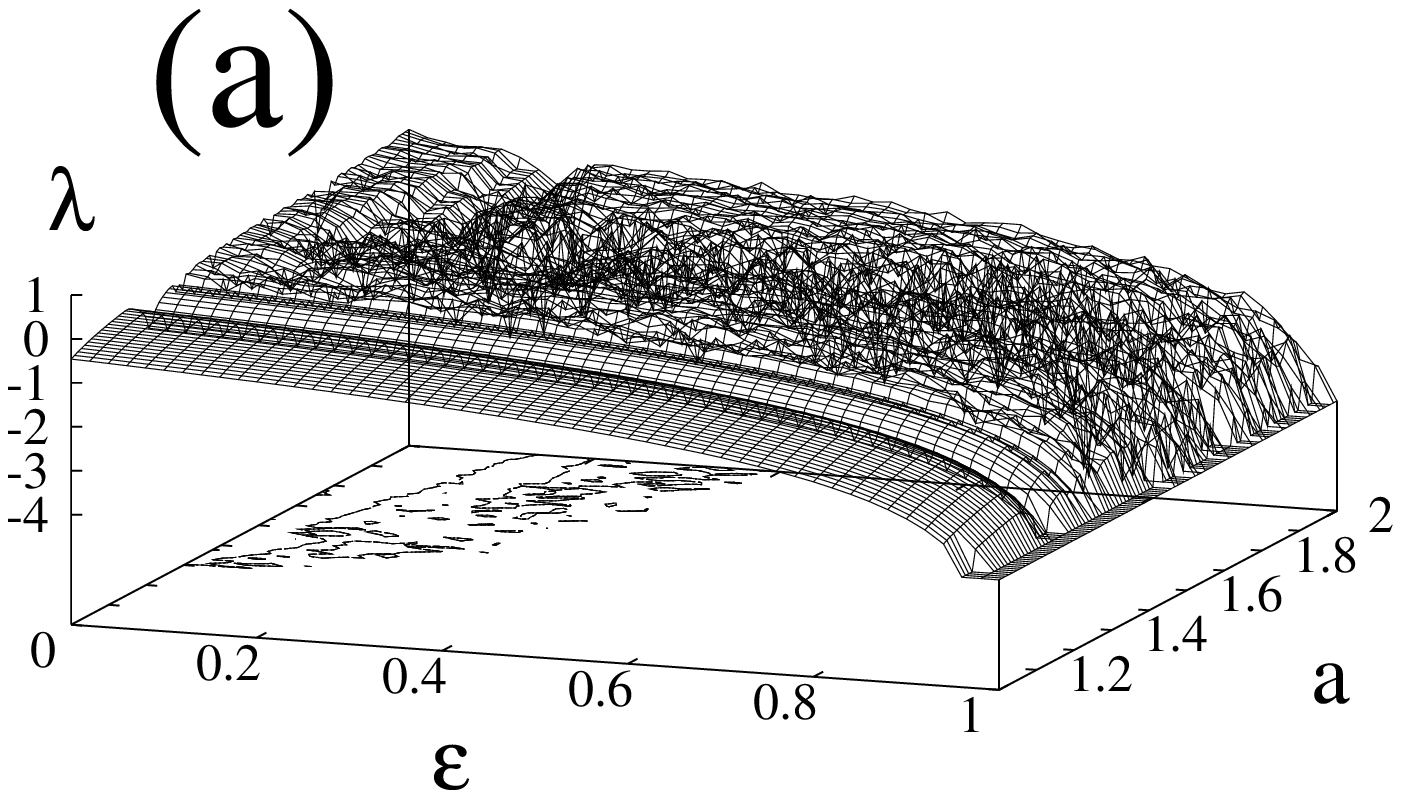}%
\includegraphics[width=8.0cm]{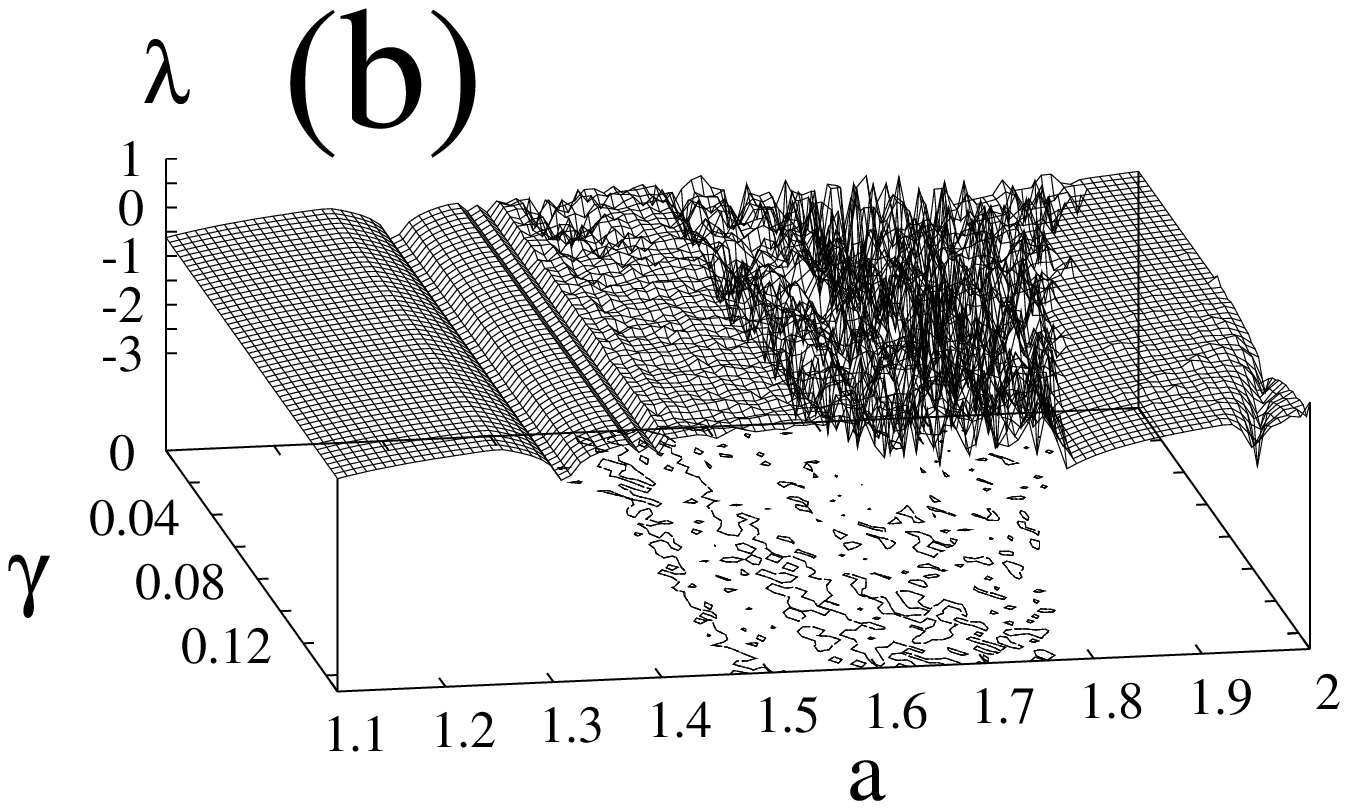}
\end{center}
\caption{\protect Variation of the Local Lyapunov spectrum as a
  function of the local nonlinearity $a$ with 
  {\bf (a)} the diffusion $\varepsilon$ for $\gamma=0$, and
  {\bf (b)} the advection $\gamma$ for $\varepsilon=0.15$.
  Here $L=64$ and contours indicate the regions where Lyapunov exponents
  are $\lambda=0$.}
\label{fig2}
\end{figure}

Figure \ref{fig2}a shows the local Lyapunov exponent as a function of both
local nonlinearity and diffusion strength.
Here we use the same conditions of Fig.~\ref{fig1} and plot $100$
values of $a$ in the range $1\le a\le 2$ for $50$ uniformly
distributed values of diffusion strength in the range $0\le\varepsilon\le 1$.
The contour lines under the figure denote loci
where local Lyapunov exponents are zero. 
From this figure one clearly sees that diffusion promotes the stability of 
local oscillators.
On the other hand, Fig.~\ref{fig2}b shows the dependence on both the local
nonlinearity and advection strength, illustrating that the local
Lyapunov spectrum does not change significantly when advection is
varied maintaining $\varepsilon$ fixed. This behavior is typical for
other values of $\varepsilon$.

As mentioned in the previous section, from local Lyapunov exponents
one cannot distinguish between absolute and convective unstable
solutions.
For negative local Lyapunov exponents there is the
possibility of having a convectively unstable pattern where
propagation of perturbations are observed~\cite{kanekobooks}. 

Next, we study the co-moving Lyapunov spectrum of pattern evolutions,
determining the co-moving Lyapunov exponents from the definition of
Deissler and Kaneko\cite{deissler} (see Ref.~\cite{kanekobooks} for details):
the logarithm of the maximum eigenvalue of the matrix $\mathbf{J}(V)$,
defined by 
\begin{eqnarray}
\mathbf{J}(V) &=& \prod_{t=1}^T
   \left [
   \begin{array}{ccc}
   \frac{\partial x_{t+1}^{1+[V(t+1)]}}
        {\partial x_t^{1+[Vt]}} 
  &\dots
  &\frac{\partial x_{t+1}^{1+[V(t+1)]}}
        {\partial x_t^{N+[Vt]}} \\
          &        &        \\
   \vdots & \ddots & \vdots \\
          &        &        \\
   \frac{\partial x_{t+1}^{N+[V(t+1)]}}
        {\partial x_t^{1+[Vt]}} 
  &\dots
  &\frac{\partial x_{t+1}^{N+[V(t+1)]}}
        {\partial x_t^{N+[Vt]}} 
   \end{array}
   \right ] \; ,\label{deissler_comov}
\end{eqnarray}
where $V$ is the velocity of the frame in which 
the Lyapunov exponent is determined, $T\gg 1$ is the number of 
time-steps used for the computation.
In Eq.~(\ref{deissler_comov}) we used the notation $x_p^q$, where $p$
represents time and $q$ represents $space$ and, in both indices
 $[z]$ represents the integer
part of $z$, and $N\le L$ represents the number of consecutive sites on 
the lattice followed by the frame.
Notice that at each time-step the matrix $\mathbf{J}(V)$ operates
between sites $\{ 1+[Vt],\dots,N+[Vt]\}$ and
sites $\{ 1+[V(t+1)],\dots,N+[V(t+1)]\}$.
For $10\le N\le 20$, we computed the matrix $\mathbf{J}(V)$ during
$T=15000$ time-steps and determined the mean value of 
of the maximum eigenvalue.

\begin{figure}[!hbp]
\begin{center}
\includegraphics[width=16.0cm]{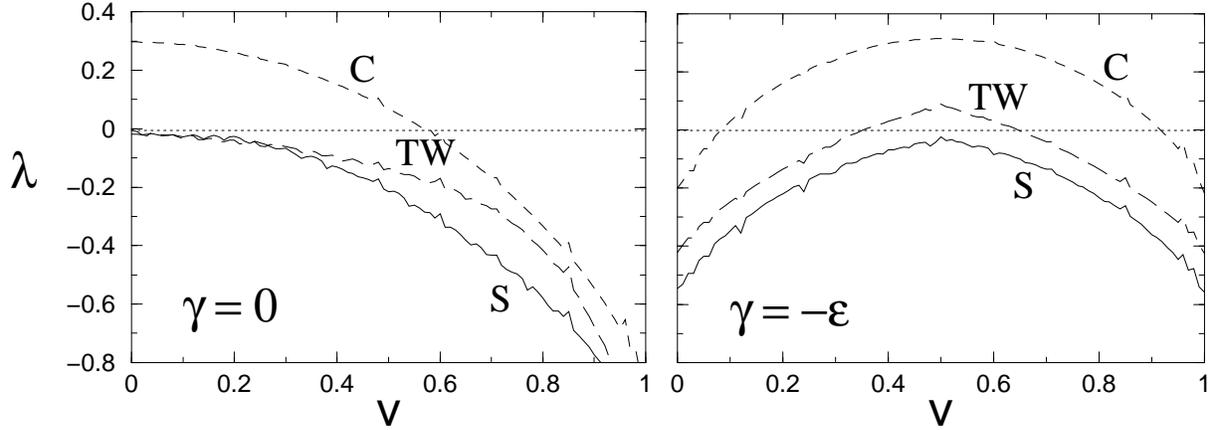}
\end{center}
\caption{\protect 
        Co-moving Lyapunov exponents, as a function of the velocity of
        the moving frame, for two extreme regimes:  
        the purely diffusive regime ($\gamma=0$) and
        the one way coupling regime ($\gamma=-\varepsilon$).
        Illustrative examples are shown, namely a chaotic pattern
        evolution ($C$) with $a=1.9$, a static pattern evolution ($S$)
        for $a=1.7$, and a traveling wave ($TW$) for $a=1.73$.
        Here $\varepsilon=0.5$ and $L=100$.}
\label{fig3}
\end{figure}

Figure \ref{fig3} shows the co-moving Lyapunov exponent for typical
examples of chaotic pattern evolutions (C) and for periodic pattern
evolutions, namely static patterns (S) and traveling waves (TW).
Two extreme regimes are illustrated: the purely diffusive regime 
($\gamma=0$) and the one-way coupling regime $\gamma=-\varepsilon$.

In the purely diffusive regime ($\gamma=0$), one clearly sees that 
chaotic pattern evolutions are absolutely unstable, 
i.e.~the local Lyapunov exponent ($V=0$) is positive, while both
periodic pattern evolutions are absolutely stable, i.e.~co-moving
Lyapunov exponents are negative for all values of the frame velocity $V$.
For one-way coupling the static pattern evolution remains absolutely 
stable while both the traveling wave and the chaotic pattern
evolution become conditionally unstable, since there is a
range of frame velocities for which one finds positive Lyapunov
exponents.

\begin{figure}[!hbp]
\begin{center}
\includegraphics[width=8.0cm]{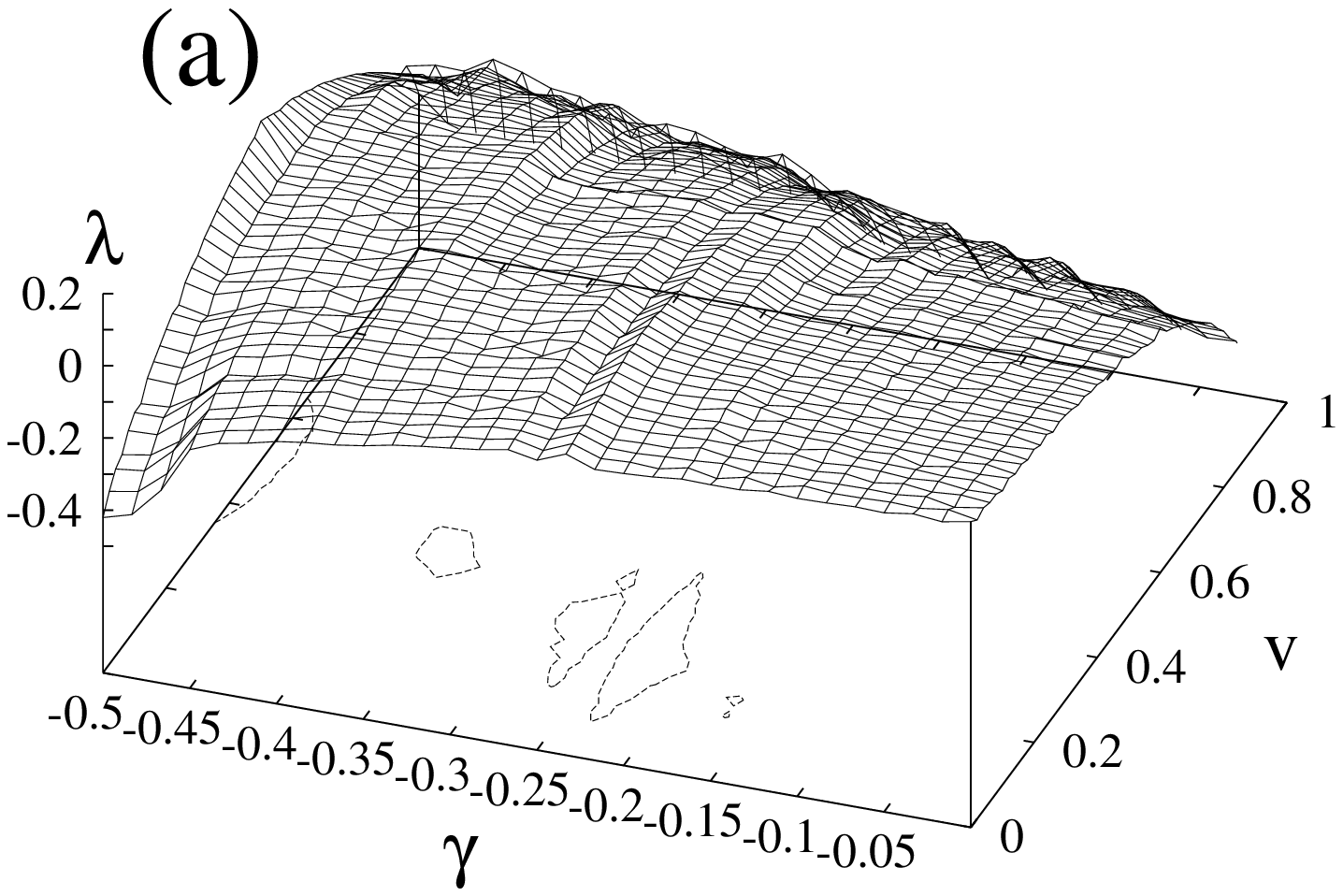}%
\includegraphics[width=8.0cm]{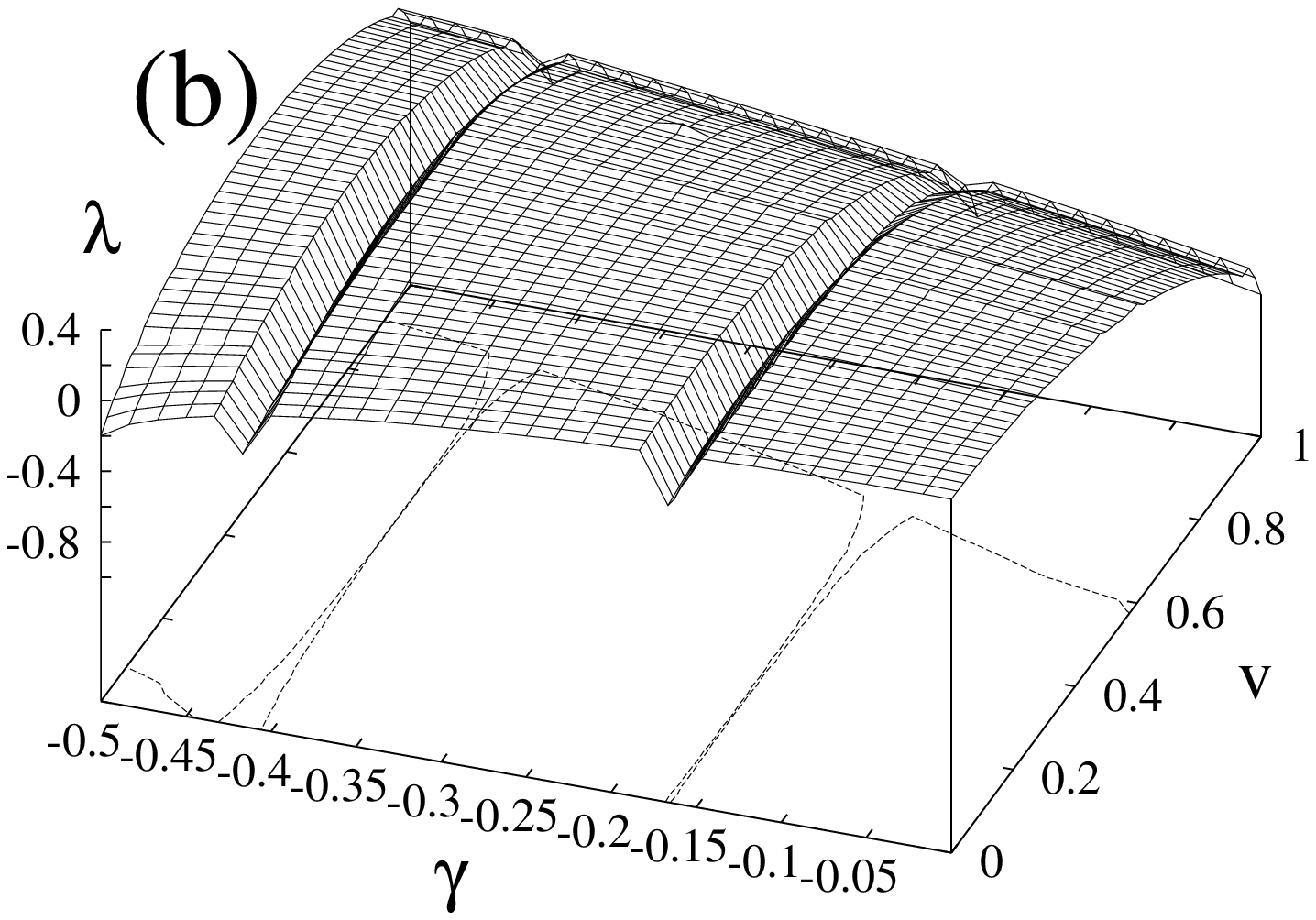}
\end{center}
\caption{\protect 
        {\bf (a)} For traveling waves, co-moving Lyapunov exponents
        increase with the advection strength, and absolutely stable
        states evolve toward convectively unstable states.
        {\bf (b)} For chaotic pattern evolutions, one observes an
        abrupt decrease of all co-moving Lyapunov exponents, when
        advection strength is increased.
        This decrease indicates that, for certain ranges of the
        advection strength, it is possible to change chaotic
        into periodic evolutions (see text).
        Here the same conditions as in Fig.~\ref{fig3} are used.}
\label{fig4}
\end{figure}

To study the transition between these two extreme regimes one needs
to compute co-moving Lyapunov exponents also as a function of the
advection $\gamma$.
Figure \ref{fig4} shows the co-moving Lyapunov exponent for
traveling waves solutions (Fig.~\ref{fig4}a) and for chaotic
pattern evolutions (Fig.~\ref{fig4}b) when advection is increased
from the purely diffusive regime to the one-way coupling regime.
The contour lines under these figures indicate the boundaries
where the co-moving Lyapunov exponent is zero.
Static evolutions are always absolutely stable when $\gamma$ varies
from $0$ to $-\varepsilon$, and therefore are not shown.

For traveling waves one observes a gradual increase of the co-moving
Lyapunov exponents reaching a convectively unstable state for very
strong advection ($\gamma\sim\vert\varepsilon\vert$), 
as shown in Fig.~\ref{fig4}a.

For chaotic pattern evolutions a surprising effect is observed:
by tuning properly the advection parameter it is possible to stabilize
unstable chaotic pattern evolutions.
In fact, from Fig.~\ref{fig4}b one clearly observes two
narrow ranges of advection strengths for which {\it absolute stability\/}
is observed, i.e.~all co-moving Lyapunov exponents become negative.
In other words, the chaotic evolution abruptly starts
to evolve periodically for these ranges of advection.
This observation strongly suggest that, in fact, {\it advection may induce
a sort of synchronization in a chaotic ensemble of nonlinear oscillators}.
Notice that the nonlinearity of the local map is strongly chaotic, 
namely one has $a=1.9$.

\section*{3. Effects of an inhomogeneous advection field}
\label{sec:heter}

In this Section we compare the results above,
obtained for uniform advection, with those for a more realistic and complicated 
situation in which the strenght and direction of advection vary 
from site to site.
To this end, for each site $i$ we choose a value
$\gamma_i=r_i\delta\varepsilon$, where $r_i$ are random numbers
homogeneously distributed 
between $-1$ and $1$ and $\delta$ measures the range where advection
strength may fluctuate, i.e. measures the `inhomogeneous
fluctuation', yielding $-\delta\varepsilon\le \gamma
\le\delta\varepsilon$, with $0\le \delta \le 1$.
For $\delta=0$ we are back to the purely diffusive regime, while
for $\delta=1$ one has maximum range of variation.

\begin{figure}[!hbp]
\begin{center}
\includegraphics[width=16.0cm]{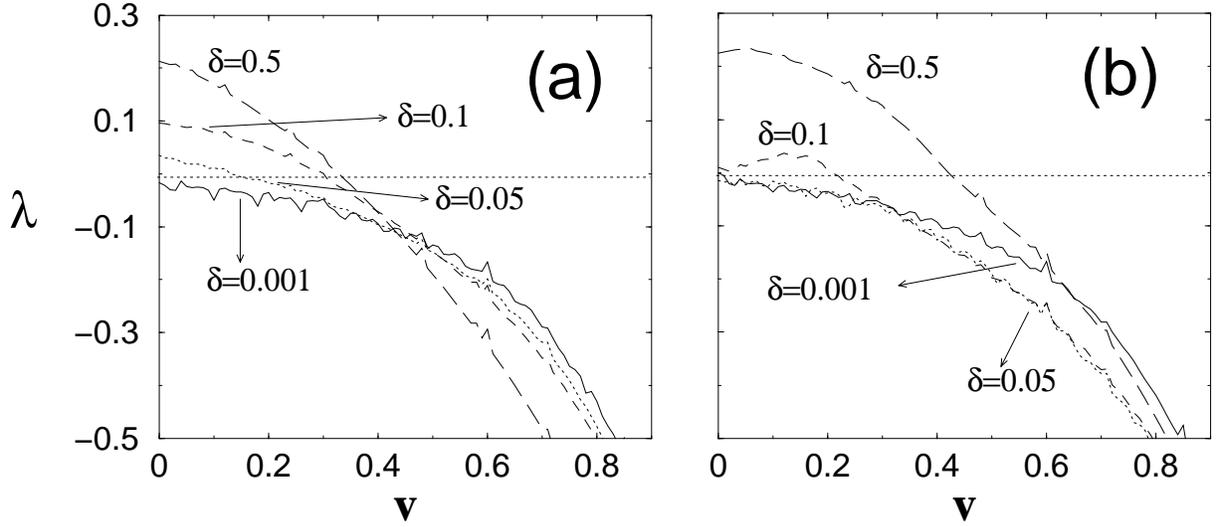}
\end{center}
\caption{\protect 
        When advection varies in space the co-moving Lyapunov
        spectrum as a function of the frame velocity increases with
        the width $\delta$ where advection $\gamma$ is randomly
        distributed (see text).
        This occurs either
        {\bf (a)} when advection direction remains unchanged, and only
        its strength is varied, and
        {\bf (b)} when both strength and direction vary randomly.
        Here we choose a traveling wave solution, for $\gamma=0$.}
\label{fig5}
\end{figure}

Figure \ref{fig5} shows the co-moving Lyapunov spectrum of a
traveling wave solution, for an inhomogeneous advection with four
different values of $\delta$, namely for $\delta=0.001,0.05,0.1,0.5$.
In Fig.~\ref{fig5}a the advection strength varies in space, but its
direction (sign) remains unchanged, namely values
of $\gamma$ are always positive, while in Fig.~\ref{fig5}b both the 
strength and the direction of advection vary randomly in space.

\begin{figure}[!hbp]
\begin{center}
\includegraphics[width=10.3cm]{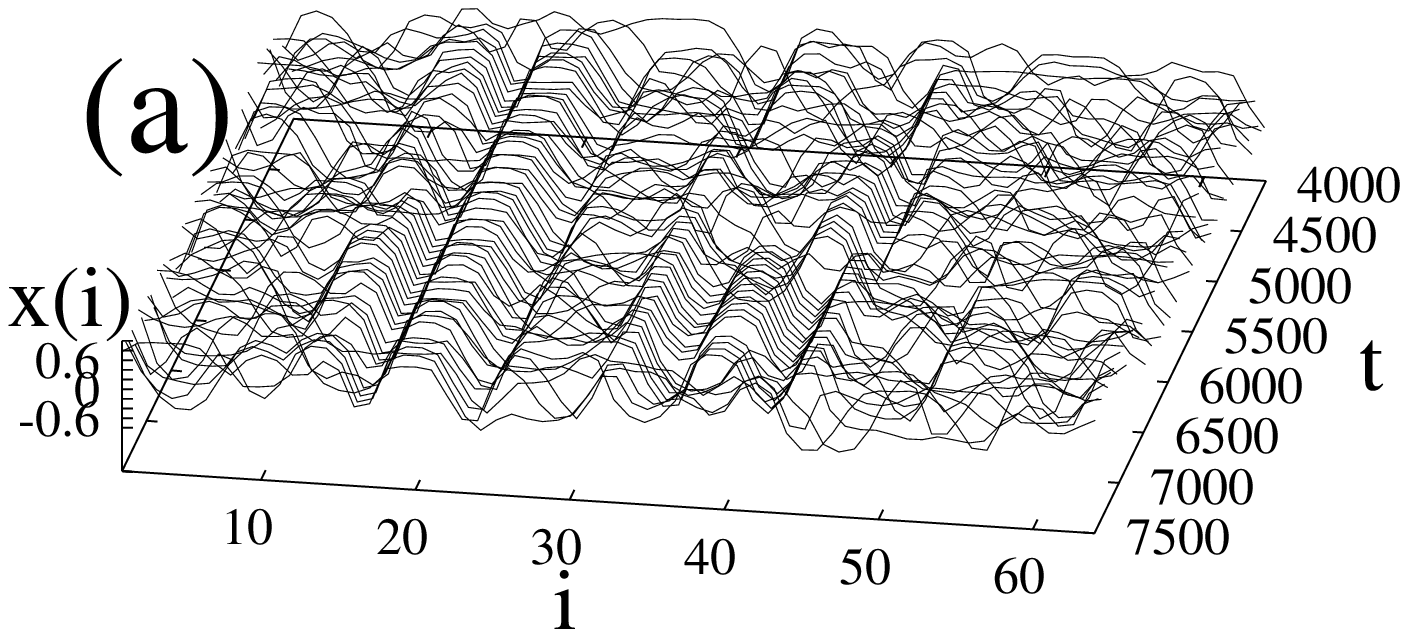}%
\includegraphics[width=5.7cm]{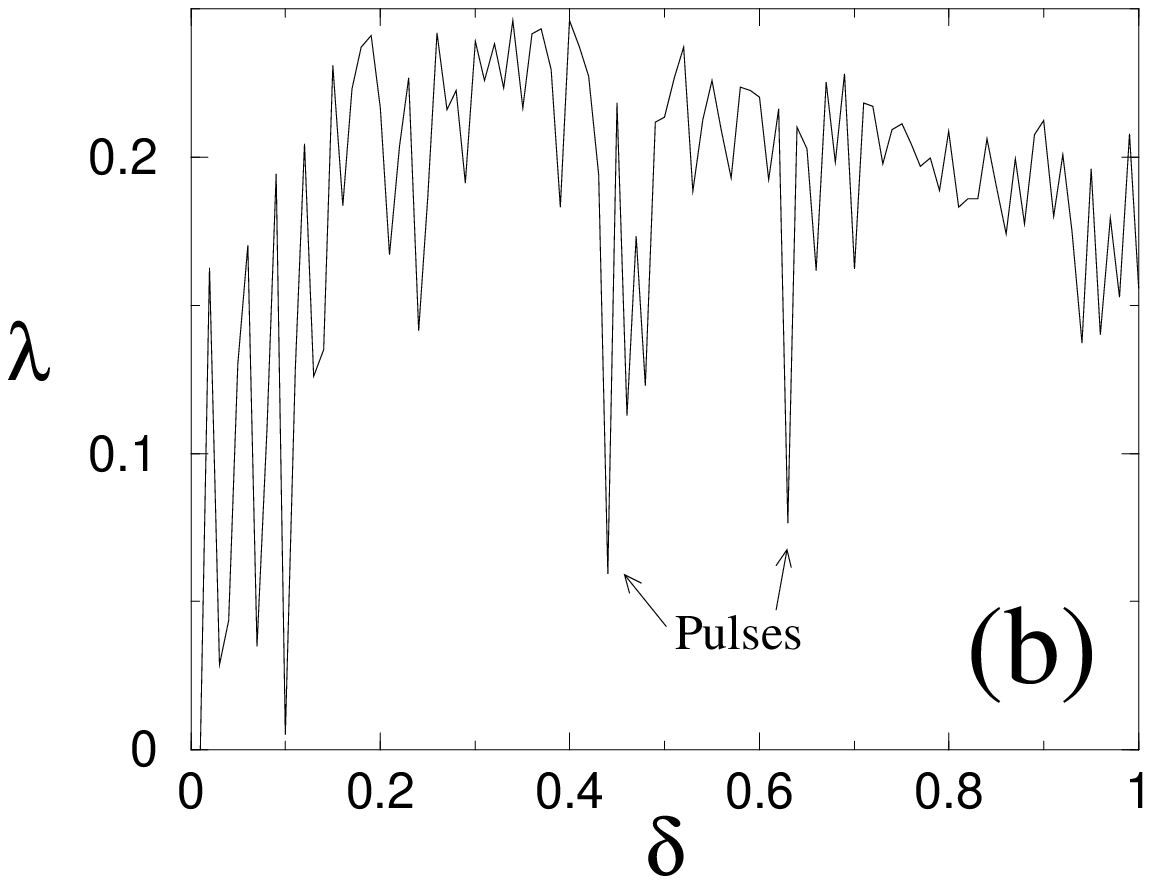}
\end{center}
\caption{\protect 
        Characterization of pulses in inhomogeneous advective
        lattices, through co-moving Lyapunov exponents. 
        {\bf (a)} Space-time diagram illustrating a typical
                  pattern evolution 
                  observed in inhomogeneous advective lattices
                  showing the appearance and disappearance of pulses
                  (see text).
        {\bf (b)} The maximum Lyapunov exponent as a function of
                  the width $\delta$ of the interval
                  $[-\delta\varepsilon,\delta\varepsilon]$ where
                  $\gamma_i$ is randomly distributed.}
\label{fig6}
\end{figure}

Comparing Fig.~\ref{fig5}a and Fig.~\ref{fig5}b
with the Lyapunov spectrum of the
traveling wave solution in Fig.~\ref{fig3}, one clearly sees that
the maximum co-moving Lyapunov exponents increases with $\delta$,
either when the direction of advection changes or not. 
As illustrated in Fig.~\ref{fig6}a,
this increase of the Lyapunov exponents is related to 
an increase of spatial disorder in the spatiotemporal
diagram of the pattern evolution.
In fact, when inhomogeneous advection is introduced, the
traveling wave cannot maintain its wave-like shape through the 
entire lattice. Instead, only certain small domains maintain
their wave-like shape. 
These domains may either remain static in an `environment' 
which evolves chaotically, or appear and disappear intermittently. 
These latter domains which appear and disappear, we call pulses.    

The occurrence of such pulses depends on the amplitude $\delta$ of the
interval where advection is varied.
Figure \ref{fig6}b shows the maximum co-moving Lyapunov exponent
as a function of $\delta$, indicating two particular values
for which pulses are observed.
As one clearly sees, pulses are characterized by an abrupt decrease of 
the maximum co-moving Lyapunov exponent, and this feature seems
to prevail when varying the set of initial conditions and the time
interval during which Lyapunov exponents are computed.

\section*{4. Conclusions}
\label{sec:conclusion}

The purpose of this manuscript is to use local and co-moving Lyapunov exponents
to characterize both periodic and aperiodic
pattern evolutions in diffusive-advective coupled map lattices as defined in
Ref.~\cite{advec}.
The effects of advection in the evolution of patterns were
studied for both homogeneous and inhomogeneous advection fields.

The main result is that there are specific ranges of advection strengths which 
induce chaotic pattern evolutions to evolve periodically. 
This is observed from the fact that for those particular advection
ranges all co-moving Lyapunov exponents turn out to be negative, 
i.e.~the pattern evolution changes from an absolutely unstable state to 
an absolutely stable state.

For the particular case of one-way coupling regime, while the static
pattern evolution remains absolutely stable, both traveling waves and
chaotic pattern evolutions become conditionally unstable.
Moreover, we showed that diffusion promotes the stability of 
local oscillators, corroborating previous studies~\cite{kanekobooks}.
For homogeneous advection, our simulations have shown that
variation of $N$ in Eq.~(\ref{deissler_comov}) does not change
significantly the final value of the co-moving Lyapunov exponent.

For an inhomogeneous advection field, traveling wave solutions are
`destroyed', and only small domains in the lattice keep their
wave-like shape. 
These domains may appear and disappear intermittently corresponding
to an abrupt decrease of the maximum co-moving Lyapunov exponent.

For inhomogeneous advection, the co-moving Lyapunov exponent
varies with the size $N$ of the matrix in Eq.~(\ref{deissler_comov}). 
In fact, preliminary results have shown that there is a pair of values 
of the matrix size $N$ and frame velocity $V$ for which the co-moving
Lyapunov exponent reaches a minimum. 
Therefore, we believe that by varying the value of $N$ one should be
able to ascertain when and where the pulses emerge in the lattice.
This interesting characterization together with  a few additional issues 
will be presented elsewhere.

\bigskip\noindent
{\large \bf Acknowledgments}

PGL thanks \textit{Funda\c{c}\~ao para a Ci\^encia e a
Tecnologia}, Portugal, for a doctoral fellowship.
JACG is a CNPq Research Fellow, Brazil.



\end{document}